\documentclass{article}

\usepackage{amssymb,nath}
\nathstyle{geometry}

\def\pd#1#2{\frac{\partial#1}{\partial#2}}
\def\eqref#1{{\rm(\ref{#1})}}

\newtheorem{proposition}{Proposition}
\newtheorem{theorem}[proposition]{Theorem}
\newtheorem{definition}[proposition]{Definition}

\let\Bbb\mathbb
\let\frak\mathfrak

\begin{document}

\title{Scalar second order evolution equations possessing an irreducible 
sl$_2$-valued zero curvature representation}

\author{Michal Marvan\thanks{Mathematical Institute, Silesian University in 
  Opava, Bezru\v{c}ovo n\'am. 13, 746 01 Opava, Czech Republic.
 {\it E-mail}: Michal.Marvan@math.slu.cz}}

\date{}

\maketitle

\begin{abstract} 
We find all scalar second order evolution equations possessing an 
sl$_2$-valued zero curvature representation that is not reducible to 
a proper subalgebra of sl$_2$.
None of these zero-curvature representations admits a parameter.
\end{abstract}

For more than twenty years, researchers are being attracted by the problem 
of classification of nonlinear systems possessing a zero-curvature 
representation (ZCR).
Efforts are focused on ZCR's taking values in a non-solvable Lie algebra 
$\frak g$ and depending on a nonremovable parameter, in expectation that 
they will be suitable for the Zakharov and Shabat~\cite{Z-S} formulation 
of integrability ({\it S-integrability}) and hence admit soliton solutions.
But even without parameter, a ZCR may be useful for construction of 
B\"acklund transformations, nonlocal symmetries and nonlocal conservation 
laws.
The problem of finding a ZCR is then equivalent to that of computing 
finite-dimensional linear coverings in the sense of Krasil'shchik and 
Vino\-gra\-dov~\cite{K-V}, which are very often just realizations
of the Wahlquist--Estabrook prolongation structures~\cite{W-E}. 
However, existing computational procedures are insufficient for solving
general classification problems, unless in combination with methods based on 
different criteria of integrability.
The most complete lists of integrable systems obtained so far resulted from 
the formal symmetry approach~\cite{M-S-S,M-S-Y,S-M}. 

In this paper we apply the method of~\cite{M1,M4} and find all second order 
scalar evolution equations
$$
u_t = F(t,x,u,u_x,u_{xx})  \numbered\label{eq}
$$
possessing an $`sl_2$-valued zero curvature representation that is 
irreducible in the sense of being not reducible to a proper subalgebra of 
$`sl_2$. 
We arrive at a single previously unnoticed class of equations parametrized 
by a single function of the coordinates $t,x$.
We also distinguish a particular subclass of equations that admit a single
conservation law. 
None of the corresponding ZCR's admits a substantial parameter, which is 
in accordance with the general belief that no second order scalar evolution 
equation is S-integrable.
All the previously known ZCR's~\cite{A-K,Rey} for second order scalar 
evolution equations turn out to be reducible to a solvable subalgebra
and as such fall outside our classification.

\section{Preliminaries}

Let $E$ be a nonlinear partial differential equation on a number of 
functions in two independent variables $t,x$. 
Let $\frak g$ be a non-solvable matrix Lie algebra.
By a {\it $\frak g$-valued zero-curvature representation} (ZCR) for $E$ 
we mean a $\frak g$-valued one-form $\alpha = A\,dx + B\,dt$ such that 
\begin{equation} \label{mc}
d\alpha = \frac12 [\alpha,\alpha]
\end{equation}
holds as a consequence of $E$.

Let $G$ be the connected and simply connected matrix Lie group associated 
with $\frak g$.
Then for an arbitrary $G$-valued function $S$, the form 
$\alpha^S = dS S^{-1} + S \alpha S^{-1}$ is another ZCR, which is said to be 
{\it gauge equivalent} to the former.
Gauge equivalent ZCR's may be viewed as identical geometric objects
(connections).
A $\frak g$-valued ZCR is said to be {\it reducible} if it is gauge 
equivalent to a ZCR taking values in a proper subalgebra of $\frak g$;
otherwise it is said to be {\it irreducible}.

Let us proceed to a description of the general algorithm of computing 
ZCR's~\cite{M1,M4} as we use it here. 
For simplicity we restrict ourselves to a single non-linear $n$th order
evolution equation
\begin{equation} \label{ev}
u_t = F(t,x,u,u_1,\dots,u_n) = 0.
\end{equation}
Here $t,x$ are coordinates, $u$ is a single field variable, and $u_1 = u_x$,
$u_2 = u_{xx}$, etc., represent the derivatives.
Let us consider the corresponding equation manifold $E$, that is, the 
infinite-dimensional space $\Bbb R^\infty$ endowed with coordinates 
$t,x,u$ and $u_k$, $1 \le k$.
The vector fields
$D_x = \pd{}{x} + u_1\,{\pd{}{u}} + \dots + u_{k+1}\,{\pd{}{u_k}} + \dots$,
$D_t = \pd{}{t} + F\,{\pd{}{u}} + \dots + D_x^kF\,{\pd{}{u_k}} + \dots$
defined on $E$ generate a diffiety structure in the sense of~\cite{K-V}
and encode all essential geometric information about the equation.
In these terms, a ZCR for eq.~\eqref{ev} is a pair of $\frak g$-valued 
functions $A,B$ on $E$ satisfying eq.~\eqref{mc}, which may be written as
\begin{equation} \label{ZCR}
D_t A - D_x B + [A,B] = 0.
\end{equation}
Let us introduce operators $\widehat D_I$ acting on an arbitrary 
$\frak g$-valued function $C$ on $E$ as follows:
\begin{equation} \label{hat}
\widehat{D}_x C = D_x C - [A,C], \quad \widehat{D}_t C = D_t C - [B,C].
\end{equation}
Operators $\widehat{D}_x,\widehat{D}_t$ commute whenever $(A,B)$ is a ZCR.
We also set $\widehat D_i
 = \widehat D_x \circ\cdots\circ \widehat D_x$ ($i$ times).

By \cite{M1} for every ZCR there exists a {\it characteristic matrix} $R$, 
which is a $\frak g$-valued function defined on $E$ (see also the independent 
work by Sakovich~\cite{Sa}).
The following proposition is proved in~\cite[Prop.~2.7 and Prop.~3.9]{M1},
see also~\cite[Prop.~2]{M4}

\begin{proposition} \label{prop}
{\rm(1)} The characteristic matrix $R$ for a ZCR of the evolution 
equation~\eqref{ev} satisfies
\begin{equation} \label{m}
-\widehat D_t R
 = \sum_i (-\widehat{D})_i (\frac{\partial F}{\partial u_i} R).
\end{equation} 

{\rm(2)} Gauge-equivalent ZCR's have conjugate characteristic matrices.
\end{proposition}

Using Proposition~\ref{prop}(2), one can restrict the gauge freedom 
by requiring that the characteristic matrix $R$ be in the Jordan normal 
form.
To fix the remaining gauge freedom, due to the stabilizer $S \subset G$ 
of~$R$, one can further transform the matrix $A$.
See Section~2 for details.

In the sequel we consider a ZCR $A\,dx + B\,dt$ taking values in $`sl_2$.
We shall write the two $`sl_2$-matrices as
\begin{equation} \label{AB}
A = (\begin{array}{cr}
a_1 & a_2 \\
a_3 & -a_1
\end{array}),\quad
B = (\begin{array}{cr}
b_1 & b_2 \\
b_3 & -b_1
\end{array}).
\end{equation}

Reducible ZCR's will be excluded from our classification. 
A subalgebra of $`sl_2$ to which the ZCR may be reduced is either an abelian 
algebra or the two-dimensional solvable subalgebra representable by 
lower triangular matrices.
Obviously, a ZCR taking values in an abelian algebra is equivalent to one
or more conservation laws.
What concerns solvable algebras, the situation is not much different.

\begin{definition} \label{excep} \rm
An $`sl_2$-valued ZCR satisfying the condition $a_2 = b_2 = 0$ is said to be 
{\it lower triangular}.
\end{definition}

For a lower triangular ZCR, it follows from eq.~\eqref{ZCR} that 
$\phi = a_1\,dx + b_1\,dt$ is a conservation law.
Let $h$ be the potential of $\phi$; then, by the same eq.~\eqref{ZCR}, 
$\phi' = (a_3\,dx + b_3\,dt)`e^{2h}$ is a conservation law nonlocal over the 
potential $h$.
This situation will be referred to as a {\it chain} of conservation laws.
Clearly, one can reconstruct the reducible ZCR from the corresponding chain 
$(\phi,\phi')$.
In this way, reducible $`sl_2$-valued ZCR's are equivalent to certain chains 
of conservation laws. 
It also follows that methods to find them are to be sought among methods to 
compute nonlocal conservation laws.

\begin{proposition} \label{a2}
Let the matrices~\eqref{AB} form a ZCR for for the evolution 
equation~\eqref{ev}.
Suppose that $a_2 = 0$. 
Then also $b_2 = 0$ or the ZCR is gauge equivalent to zero.
\end{proposition}

\paragraph*{Proof}
Let us denote by $C$ the matrix~\eqref{ZCR} evaluated at $a_2 = 0$.
By assumption, $C$ is zero on the equation manifold $E$.
If $b_2 \ne 0$, then from the condition $0 = c_2 = -D_x b_2 + 2 a_1 b_2$ 
we compute that $a_1 = \frac12 D_x b_2/b_2$ on $E$, and then from the 
condition $0 = c_1 = D_t a_1 - D_x b_1 - a_3 b_2$ we compute that
$a_3 = \frac12 \frac{D_{tx}b_2}{b_2^2}
 - \frac12 \frac{D_t b_2 D_x b_2}{b_2^3}
 - \frac{D_x b_1}{b_2}
$ on $E$.
Let us introduce a function $G$ by the requirement
$b_3 = -\frac{b_1^2}{b_2}
 - \frac{D_t b_1}{b_2}
 + \frac{b_1 D_t b_2}{b_2^2}
 + \frac12 \frac{D_{tt}b_2}{b_2^2}
 - \frac34 \frac{(D_t b_2)^2}{b_2^3}
 + \frac{G}{b_2}$.
Then $0 = c_3 = -D_x G/b_2$, which in the case of an evolution equation 
implies that $G$ is a function of $t$ only.
Then, under the above substitutions for $a_1,a_3,b_3$, the gauge matrix
$$
(\begin{array}{cc}
b_2^{-1/2} & 0 \\ -\frac12  D_t b_2 b_2^{-3/2} + b_1 b_2^{-1/2} & b_2^{1/2}
\end{array})
$$
sends $A$ to zero and $B$ to 
$$
(\begin{array}{cc}
0 & 1 \\ G & 0
\end{array}),
$$
which depends on $t$ at most.
The last matrix is sent to zero by gauge transformation with the gauge 
matrix composed of independent solutions of the equation $s_{tt} = G s$.

\section{The classification}

Let us consider a second order evolution equation~\eqref{eq}
along with the $`sl_2$-matrices $A,B$ satisfying eq.~\eqref{ZCR} but not
reducible to a solvable subalgebra.
We also assume that $\frac{\partial F}{\partial u_{xx}} \ne 0$.
Following~\cite{M4}, we consider the two cases distinguished by 
their Segre characteristics separately.

\subsection{The nilpotent case}

% File Evol2 N

Under the notation~\eqref{AB}, the Jordan form for $R$ corresponds to 
$r_1 = 0$, $r_2 = 0$, $r_3 = 1$.
The normal form for $A$, obtained in~\cite{M4}, is given by the single 
requirement $a_1 = 0$.
Indeed, whenever $a_2 \ne 0$ (otherwise the ZCR is either lower triangular 
or trivial by Proposition~\ref{a2}), then one can set $a_1 = 0$ in a general 
matrix $A$ by means of the gauge matrix 
$$
(\begin{array}{cc}
1 & 0 \\ a_1/a_2 & 1
\end{array})
$$
from the stabilizer of $R$.

The equation \eqref{m} then reduces to the system $T_i = 0$, $i = 1,2,3$, 
where
\begin{eqnarray*}
T_1 &:=& 2 D_x (\pd F{u_{xx}}) a_2 + \pd F{u_{xx}} D_x a_2 - \pd F{u_x} a_2
 + b_2,
\\
T_2 &:=& 2 \pd F {u_{xx}} a_2^2,
\\
T_3 &:=& - D_{xx}\pd F{u_{xx}} + D_x \pd F{u_x} - 2 \pd F{u_{xx}} a_2 a_3
 - \pd F u - 2 b_1.
\end{eqnarray*}
Then $a_2 = 0$ by the second equation and $b_2 = 0$ by the first equation,
whence the ZCR is lower triangular.
Consequently, this case is void in our classification.

\subsection{The semisimple case}

% File Evol2 SS max

It will be convenient to change the notation for $`sl_2$-matrices to
$$
A = (\begin{array}{cc}
a_1 & a_2 + a_3 \\
a_2 - a_3 & \llap{$-$}a_1
\end{array}).
$$
The Jordan form for $R$ has $r_2 = r_3 = 0$ with $r := r_1$ arbitrary.
Unlike in~\cite{M4}, we choose the normal form for~$A$ characterized by 
the single requirement $a_3 = 0$.
And indeed, whenever $a_2 + a_3 \ne 0$, which is irrestrictive by 
Proposition~\ref{a2}, one can set $a_3$ to zero by a gauge transformation
from the stabilizer of $R$. 
The relevant gauge matrix is diagonal with the diagonal entries $h$ and 
$1/h$, where $h = (\frac{a_2 - a_3}{a_2 + a_3})^{1/4}$.

Eq. \eqref{ZCR} and \eqref{m} then assume the form $S_i = 0 = T_i$, 
$i = 1,2,3$, with
\begin{equation} \label{SS}
\wall
S_1 = -D_t a_1 + D_x b_1 + 2 a_2 b_3,
\\
S_2 = -D_t a_2 + D_x b_2 - 2 a_1 b_3,
\\
S_3 = D_x b_3 + 2 a_2 b_1 - 2 a_1 b_2,
\\
T_1 = -D_t r
 - \frac {\partial F}{\partial u} r
 + D_x(\frac {\partial F}{\partial u_x} r)
 - D_{xx}(\frac {\partial F}{\partial u_{xx}} r) 
 - 4 \frac {\partial F}{\partial u_{xx}} ra_2^2, 
\\
T_2 = 
-b_3 + 2 \frac {\partial F}{\partial u_{xx}} a_1 a_2, 
\\
T_3 = -b_2
 + \frac {\partial F}{\partial u_x} a_2
 - 2 D_x (\frac {\partial F}{\partial u_{xx}}) a_2
 - 2{\frac{D_x r}{r}} {\frac {\partial F}{\partial u_{xx}}} a_2
 - \frac {\partial F}{\partial u_{xx}
} D_x a_2. 
\return
\end{equation}

If $a_2 = 0$, then we have $b_2 = b_3 = 0$ by the last two equations and 
the ZCR reduces to a single conservation law.
Therefore, we assume that $a_2 \ne 0$ in the sequel.

\begin{proposition} \label{prop1}
As solutions to eq. \eqref{SS}, functions $r,a_1,a_2,b_3$ cannot depend 
on jet coordinates other than $t,x,u,u_x,u_{xx}$,
whereas functions $b_1,b_2$ cannot depend on jet coordinates other than 
$t,x,u,u_x,u_{xx},u_{xxx}$.
\end{proposition} 

\paragraph*{Proof}
We may assume, without loss of generality, that the functions 
$r,a_1,\ a_2,b_3$ 
depend on $t,x,\ u,\dots,u_k$ and the functions $b_1,b_2$ depend on 
$t,x,\ u,\dots,\ u_k,u_{k+1}$ for some $k \ge 2$.
We perform a downward induction, each step of which consists in deriving
appropriate differential consequences of the system~(\ref{SS}). 
Thus, let $k \gt 2$. Then we have
$$
0 = \frac{\partial T_1}{\partial u_{k+2}}
 = -2 \frac{\partial F}{\partial u_{xx}} \frac{\partial r}{\partial u_k},
$$
but $\frac{\partial F}{\partial u_{xx}} \ne 0$,
whence $r$ does not depend on $u_k$.
Then, similarly,
$$
0 = \pd{S_2}{u_{k+2}} - \pd{T_3}{u_{k+1}}
 = 2 \frac{\partial b_2}{\partial u_{k+1}} 
\\
0 = \pd{S_2}{u_{k+2}} + \pd{T_3}{u_{k+1}}
 = -2 \frac{\partial F}{\partial u_{xx}} \frac{\partial a_2}{\partial u_k}, 
$$
whence $b_2$ does not depend on $u_{k+1}$ and $a_2$ does not depend on $u_k$. 
Finally,
$$
0 = -2 a_2 \pd{S_1}{u_{k+2}} + \pd{S_3}{u_{k+1}} + \pd{T_2}{u_k}
 = 4 a_2 \frac{\partial F}{\partial u_{xx}} \frac{\partial a_1}{\partial u_k}, 
\\
0 = \phantom{-}2 a_2 \pd{S_1}{u_{k+2}} + \pd{S_3}{u_{k+1}} + \pd{T_2}{u_k} 
 = 4 a_2 \frac{\partial b_1}{\partial u_{k+1}},
\\
0 = -2 a_2 \pd{S_1}{u_{k+2}} + \pd{S_3}{u_{k+1}} - \pd{T_2}{u_k}  
 = 2 \frac{\partial b_3}{\partial u_k}, 
$$
whence
$a_1$, $b_3$ do not depend on $u_k$ and
$b_1$ does not depend on $u_{k+1}$ (recall that $a_2 \ne 0$).
This completes the induction step.
\bigskip

Under the restrictions established in Proposition~\ref{prop1}, the 
determining system (\ref{SS}) becomes an overdetermined system of partial 
differential equations. 
As such, it can be solved routinely, but its solution is troublesome even 
with the employment of software capable of automated deriving of
differential consequences.
The reason is that the class of second order evolution equations 
is invariant with respect to a large group of contact transformations
$\bar x = \bar x(t,x,u,u_x)$, 
$\bar u = \bar u(t,x,u,u_x)$, 
$\bar t = \bar t(t)$.
Below we shall apply a series of suitably chosen contact transformations 
to achieve substantial reduction of the matrix~$A$.

% File Evol2 SS j2

\begin{proposition} \label{propj2}
For every second order evolution equation {\rm(\ref{eq})} possessing an 
irreducible $`sl_2$-valued ZCR there exists a contact transformation 
such that the transformed $a_2$ depends on $t,x,u,u_x$ at most.
\end{proposition}

\paragraph*{Proof}
Let functions $r,a_i,b_i$ depend on $t,x,u,u_x,u_{xx},u_{xxx}$ as 
in~Proposition~\ref{prop1}.
Taking successively the derivatives
$\pd{S_1}{u_{xxxx}}$,
$\pd{S_2}{u_{xxxx}}$,
$T_2$,
$\pd{T_3}{u_{xxx}}$,
$\pd{T_1}{u_{xxxx}}$,
$\pd{S_3}{u_{xxx}}$,
$\pd{T_3}{u_{xxx}}$,
$\pd{^2 S_1}{u_{xxx}^2}$,
$\pd{^2 S_2}{u_{xxx}^2}$
one may check routinely that
$$
\frac{\partial^2 a_2}{\partial u_{xx}^2} = 0
\quad\text{and}\quad
a_1 \frac{\partial a_2}{\partial u_{xx}}
  - \frac{\partial a_1}{\partial u_{xx}} a_2 = 0,
$$
are among differential consequences of the system (\ref{SS}).
Hence, 
\begin{enumerate}
\item[(a)] $a_2$ is linear in $u_{xx}$, i.e.,
$a_2 = a_{21}(t,x,u,u_x) u_{xx} + a_{20}(t,x,u,u_x)$;
\item[(b)] the ratio $a_1/a_2$ does not depend on $u_{xx}$.
\end{enumerate}
Now, if $a_{21} = 0$, then the statement is proved.
Otherwise, let $f_1,f_2$ be two functionally independent solutions of the 
linear equation
$$ \numbered \label{f}
-\frac{a_{20}}{a_{21}} \pd f{u_x} + u_x \pd f u + \pd f x = 0.
$$
In particular, both $f_1$ and $f_2$ do depend on $u_x$.
Then $\bar t = t$, $\bar x = f_1$, $\bar u = f_2$ and 
$\bar u_{\bar x} = \frac{\pd{f_2}{u_x}}{\pd{f_1}{u_x}}$ 
satisfy the well-known necessary conditions of being a contact 
transformation:
$$
\frac{\pd{\bar u}{u_x}}{\pd{\bar x}{u_x}}
 = \bar u_{\bar x}
 = \frac{\pd{\bar u}{x} + u_x \pd{\bar u}{u}}
   {\pd{\bar x}{x} + u_x \pd{\bar x}{u}}.
$$
Under this transformation, $A\,dx + B\,dt$ becomes 
$\bar A\,d\bar x + \bar B\,d\bar t$ 
with $d\bar x = D_x \bar x\,dx + D_t \bar x\,dt$, $d\bar t = dt$, so that
$$
A = \bar A\,D_x \bar x
 = (\pd{f_2}{x} + u_x \pd{f_2}{u} + u_{xx} \pd{f_2}{u_x}) \bar A
 = \frac{a_2}{a_{21}} \pd{f_2}{u_x} \bar A,
$$
where we have used eq.~\eqref{f}.
Hence
$$
\bar A = \frac{a_{21}}{\partial f_2/\partial u_x} 
 (\begin{array}{c@{\quad}c}
   a_1/a_2 & 1 \\ 1 & \llap{$-$}a_1/a_2 
  \end{array}),
$$
which is independent of $u_{xx}$, hence of $\bar u_{xx}$, by virtue of 
statement~(b) above.

\begin{theorem} \label{theorem}
Every second order scalar evolution equation {\rm(\ref{eq})} possessing an 
irreducible $`sl_2$-valued ZCR is transformable to an equation of the 
form
\begin{equation} \label{result}
u_t = 
\frac{\partial\beta}{\partial x} u^2 u_{xx}
 + 2 \frac{\partial^2\beta}{\partial x^2} u^2 u_x
 + 4 \beta u_x
 + (\frac{\partial^3\beta}{\partial x^3}
 - 4 \frac{\partial\beta}{\partial x}) u^3
 - 4 \frac{\partial\beta}{\partial x} u
\end{equation}
through a contact transformation.
Here $\beta$ is an arbitrary function of $t,x$ with $\pd \beta x \ne 0$.
The ZCR is then $A\,dx + B\,dt$ with
$$ \numbered \label{result zcr}
A = (\begin{array}{c@{\quad}c} \frac 1u & 1 \\ 1 & \llap{$-$}\frac 1u 
\end{array}),
\\
B = (\begin{array}{cc} 
-\frac{\partial\beta}{\partial x} u_x + 4 \frac \beta u
 - \frac{\partial^2 \beta}{\partial x^2} u
& 
4 \beta + 2 \frac{\partial\beta}{\partial x} u
\\
4 \beta - 2 \frac{\partial \beta}{\partial x} u
& 
\frac{\partial\beta}{\partial x} u_x - 4 \frac \beta u
 + \frac{\partial^2 \beta}{\partial x^2} u
\end{array}).
$$
\end{theorem}

\paragraph*{Proof}
Following Proposition~\ref{propj2}, we assume that the matrix $A$ depends 
on $t,x,u,u_x$ at most.
One may check routinely that
$$
\pd{^2 a_2}{u_x^2} = 0
\quad\text{and}\quad
\pd{^2 a_2}{x\,\partial u_x}
 + u_x \pd{^2 a_2}{u\,\partial u_x}
 - \pd{a_2}{u} = 0
$$
are among differential consequences of the system~\eqref{SS}.
The general solution is
$a_2 = \pd h x + u_x\,{\pd h u} = D_x h$ for a suitable function $h(t,x,u)$.
If $a_2$ does depend on $u_x$, then $\pd h u \ne 0$, whence
$\bar t = t$, $\bar x = h$, $\bar u = x$ is a point transformation.
If $a_2$ does not depend on $u_x$, then $h$ does not depend on $u$,
but does depend on $x$ (otherwise $a_2 = 0$), and $\bar t = t$, 
$\bar x = h$, $\bar u = u$ is a point transformation.
In both cases $A = \bar A\,D_x h = \bar A a_2$, 
whence $\bar a_2 = 1$ in the transformed matrix $\bar A$.

With $a_2 = 1$, one can check routinely that $\partial a_1/\partial u_x = 0$ 
is among differential consequences of system~\eqref{SS}.
% File reduction f(t,x,u,u_x) %
If moreover $\partial a_1/\partial u = 0$, then $A$ is completely independent 
of $u$ and its derivatives, and then so is $B$, whence the 
ZCR is gauge equivalent to zero.
Therefore, we shall continue with $\partial a_1/\partial u \ne 0$.
Then we can apply a point transformation $\hat x = \bar x$, 
$\hat u = 1/a_1$, which sends $a_1$ to $1/\hat u$
(this choice prevents terms quadratic in $u_x$ from appearing on the 
right-hand side of eq.~\eqref{eq}).
It is then a matter of routine to compute all possible forms of the
right-hand side $F$ of eq.~\eqref{eq} and also the corresponding 
matrices~$B$.
\bigskip

There seem to be no earlier appearance of the class~\eqref{result} in the
literature, let alone its `simplest' member 
$u_t = u^2 u_{xx} + 4 x u_x - 4 u^3 - 4 u$.

The results would be incomplete if we do not establish irreducibility of the 
ZCR~\eqref{result zcr}.
Since reducibility implies existence of at least one local conservation law, 
we shall start with the following result.

\begin{proposition}
Within the class~\eqref{result}, the only equations to possess a conservation 
law are those with
$$ \numbered \label{beta}
\beta = \frac18
\frac {p_t `e^{2x} + q_t `e^{-2x}}{p `e^{2x} - q `e^{-2x}},
$$
where $p,q$ are arbitrary functions of $t$ such that $(pq)_t \ne 0$.
In all these cases the equation has a single conservation law 
$$ \numbered \label{beta cl}
D_t \frac{p `e^{2x} + q `e^{-2x}}{u}
= D_x(\wall
   \frac12 
   \frac{(pq)_t(p`e^{2x} + q`e^{-2x})}
        {(p`e^{2x} - q`e^{-2x})^2}
   u_x
\\ +
   \frac12 
   \frac{(p_t`e^{2x} + q_t`e^{-2x})(p`e^{2x} + q`e^{-2x})}
       {(p`e^{2x} - q`e^{-2x})}
   \frac 1u
\\ -
   \frac{(pq)_t (3 p^2 `e^{4x} + 2 p q + 3 q^2 `e^{-4x})}
        {(p`e^{2x} - q`e^{-2x})^3}
   u
  ).
  \return
$$
\end{proposition}

\paragraph*{Proof}
A routine computation shows that any characteristics $\psi$ of a 
conservation law depends on $t,x,u$ at most and satisfies the equations
$$
\frac{\partial^2 \psi}{\partial x^2} - 4 \psi = 0,
\quad
\frac{\partial \psi}{\partial u} + 2 \frac \psi u = 0,
\quad 
\frac{\partial \psi}{\partial t} - 4\beta \frac{\partial \psi}{\partial x}
 = 0.
$$
The rest is easy.
\bigskip

Another computation shows that for none of the equations of the 
class~\eqref{beta} the corresponding ZCR~\eqref{result zcr} can be 
reduced to the lower triangular form with multiples of~\eqref{beta cl} 
on the diagonal.
Thus, the ZCR's~\eqref{result zcr} are indeed irreducible.

Finally, a remark on equations determining pseudospherical surfaces 
(PSS equations) is due.
In anticipation of finding new S-integrable nonlinear systems, a number of 
attempts have been made to classify equations describing pseudospherical 
surfaces (PSS equations), see~\cite{T} and references therein.
Even though being a PSS equation is equivalent to possessing an 
$`sl_2$-valued ZCR, the classification of second order scalar evolution 
PSS equations as obtained by Reyes~\cite{Rey} (see also~\cite{Rey',F-O-R}) 
has no intersection with ours. 
This seeming paradox is easily resolved.
In fact, each of the ZCR's found by Reyes is reducible to the lower 
triangular form (the generalized Burgers equation) or even to a single 
conservation law (the other equations), which are disregarded in our 
classification.
On the other side, equations~\eqref{result} are not integrable, hence 
do not enter the classification of integrable equations by Svinolupov and 
Sokolov~\cite{Sv,S-S}, which was the starting point of the Reyes work.

\section*{Acknowledgements}

The support from the grants MSM:J10/98:192400002 and \endgraf\noindent
GA\v{C}R 201/00/0724 is gratefully acknowledged.


\small

\begin{thebibliography}{99}

\bibitem{A-K}
A.A. Alexeyev and N.A. Kudryashov,
Non-Abelian pseudopotentials and conservation laws of 
reaction-diffusion equations, {\it J. Phys. A: Math. Gen.} {\bf 24}
(1991) L255--L260.

\bibitem{F-O-R} 
M.V. Foursov, P.J. Olver and E.G. Reyes, On formal integrability of evolution
equations and local geometry of surfaces, {\it Diff. Geom. Appl.} {\bf 15} 
(2001) 183--199.

\bibitem{K-V} 
I.S. Krasil'shchik and A.M. Vinogradov,
Nonlocal trends in the geometry of differential equations: symmetries,
conservation laws, and B\"acklund transformations,
{\it Acta Appl. Math.} {\bf 15} (1989) 161--209.

\bibitem{M1}
M. Marvan, 
On zero curvature representations of partial differential equations, 
in: {\it Differential Geometry and Its Applications},
Proc. Conf. Opava, Czechoslovakia, Aug. 24--28, 1992 
(Silesian University, Opava, 1993) 
103--122.
Electronic version in ELibEMS at http://www.emis.de/proceedings.

\bibitem{M4}
M. Marvan, 
A direct procedure to compute zero-curvature 
representations. The case sl$_2$, in: {\it Secondary Calculus
and Cohomological Physics,} Proc. Conf. Moscow, 1997 (ELibEMS, 
http://www.emis.de/proceedings/SCCP97, 1998) pp.10.

\bibitem{M-S-S}
A.V. Mikhailov, A.B. Shabat and V.V. Sokolov,
The symmetry approach to classification of integrable equations,
in: V.E. Zakharov, ed., {\sl What is Integrability?}
(Springer, Berlin, 1991) 115--184.

\bibitem{M-S-Y}
A.V.~Mikhailov, A.B.~Shabat and R.I.~Yamilov,
Extension of the module of invertible transformations. 
Classification of integrable systems,
{\it Commun. Math. Phys.} {\bf 115} (1988) 1--19.

\bibitem{Rey}
E.G. Reyes, Pseudo-spherical surfaces and integrability of evolution
equations, {\it J. Differential Equations} {\bf 147} (1998) 195--230.

\bibitem{Rey'}
E.G. Reyes, Some geometric aspects of integrability of differential 
equations in two independent variables, {\it Acta Appl. Math.} 
{\bf 64} (2000) 75--109.

\bibitem{S-M} 
A.B.~Shabat and A.V.~Mikhailov, Symmetries---test of integrability, 
in: A.S. Fokas and V.E. Zakharov, eds., {\it Important Developments 
in Soliton Theory} (Springer, Berlin, 1993) 355--374.

\bibitem{Sa}
S.Yu.~Sakovich,
On zero-curvature representations of evolution equations,
{\it J. Phys. A: Math. Gen.} {\bf 28} (1995) 2861--2869.

\bibitem{Sv}
S.I. Svinolupov, Second-order evolution equations with symmetry,
{\it Uspekhi Matem. Nauk} {\bf 40} (1985) (5) 263--264 (in Russian).

\bibitem{S-S}
S.I. Svinolupov and V.V. Sokolov, Weak nonlocalities in evolution 
equations, {\it Mat. Zametki} {\bf 48} (1990) (6) 91--97 (in Russian); 
{\it Math. Notes} {\bf 48} (1990) (5--6) 1234--1239.

\bibitem{T} 
K. Tenenblat, 
{\it Transformations of Manifolds and Applications to Differential 
Equations,}
Pitman Monographs and Surveys in Pure and Applied Mathematics 93
(Longman, London, 1998).
  
\bibitem{W-E} 
H.D. Wahlquist and F.B. Estabrook,
Prolongation structures and nonlinear evolution equations I, II,
{\it J. Math. Phys.} {\bf 16} (1975) 1--7; {\bf 17} (1976) 1293--1297.

\bibitem{Z-S}
V.E. Zakharov and A.B. Shabat,
Integration of nonlinear equations of mathematical physics by the method of 
inverse scattering. II, {\it Funct. Anal. Appl.} {\bf 13} (1980) 166--174; 
translation from {\it Funkc. Anal. Prilozh.} {\bf 13} (1979) (3) 13--22.



\end{thebibliography}
\end{document}